\documentclass[epj,twocolumn]{webofc}
\usepackage[varg]{txfonts}   % Web of Conferences font
\usepackage{epstopdf}
\def\cC{{\cal C}}
\def\cL{{\cal L}}
\def\cO{{\cal O}}
\def\cR{{\cal R}}
\def\kW{\kappa_{\raisebox{-.03cm}{\scriptsize $W$}}}
\def\kV{\kappa_{\raisebox{-.03cm}{\scriptsize $V$}}}
\def\kF{\kappa_{\raisebox{-.03cm}{\scriptsize $F$}}}
\def\kt{\kappa_{\raisebox{-.02cm}{\scriptsize $t$}}}
\def\ku{\kappa_{\raisebox{-.02cm}{\scriptsize $u$}}}
\def\kf{\kappa_{\raisebox{-.02cm}{\scriptsize $f$}}}
\woctitle{LHCP 2013}
\begin{document}
%
%\title{Higgs Physics}
\title{The Physics of the Higgs-like Boson}

\author{Antonio Pich\inst{1}\fnsep\thanks{\email{Antonio.Pich@ific.uv.es}}
}

\institute{Departament de F\'\i sica Te\`orica, IFIC, Universitat de Val\`encia -- CSIC, Apt. Correus 22085, 46071 Val\`encia, Spain}

\abstract{The present knowledge on the Higgs-like boson discovered at the LHC is summarized. The data accumulated so far are consistent with the Standard Model predictions and put interesting constraints on alternative scenarios of electroweak symmetry breaking.
The measured couplings to gauge bosons and third-generation fermions indicate that a Higgs particle has indeed been found. More precise data are needed to clarify whether it is the unique Higgs boson of the Standard Model or the first member of a new variety of
dynamical (either elementary or composite) fields.}
\maketitle
\section{Introduction}
\label{sec:introduction}

The data accumulated so far  \cite{Aad:2012tfa,Aad:2013wqa,Chatrchyan:2012ufa,Chatrchyan:2013lba,Aaltonen:2012qt} confirm the Higgs-like nature \cite{Higgs:1964pj,Englert:1964et,Guralnik:1964eu,Kibble:1967sv} of the new boson discovered at the LHC, with a spin/parity consistent with the Standard Model (SM) $0^+$ assignment \cite{Aad:2013xqa,Chatrchyan:2012jja,D0-6387}. The observation of its $2\gamma$ decay mode demonstrates that it is a boson with $J\not=1$, while the $J^P=0^-$ and $2^+$ hypotheses have been already excluded at confidence levels above 99\%,  analysing the distribution of its decay products.
The masses measured by ATLAS ($M_H = 125.5 \pm 0.2 \,{}^{+\, 0.5}_{-\, 0.6}$~GeV) and CMS ($M_H = 125.7 \pm 0.3 \pm 0.3$~GeV) are in good agreement, giving the average value
\begin{equation}\label{eq:Higgs_Mass}
M_H = 125.64 \pm 0.35~\mathrm{GeV}\, .
\end{equation}

Although its properties are not well measured yet, it complies with the expected behaviour and, therefore, it is a very compelling candidate to be the SM Higgs.
An obvious question to address is whether it corresponds to the unique Higgs boson incorporated in the SM, or it is just the first signal of a much richer scenario
of Electroweak Symmetry Breaking (EWSB). Obvious possibilities are an extended scalar sector with additional fields or dynamical (non-perturbative) EWSB generated by some new underlying dynamics.
While more experimental analyses are needed to assess the actual nature of this boson, the
present data give already very important clues, constraining its couplings in a quite significant way.

Whatever the answer turns out to be, the LHC findings represent a truly fundamental discovery with far reaching implications. If it is an elementary scalar (the first one), one would have established the existence of a bosonic field (interaction) which is not a gauge force. If it is instead a composite object, there should be a completely new underlying interaction.

\section{Standard Model Higgs Mechanism}
\label{sec:SM}

%%%%%%%%%%%%%%%%%%%%%% Figure Higgs Potential %%%%%%%%%%%%%%%%%%%%%
\begin{figure}
%\centering
\begin{center}
\includegraphics[width=4.5cm,clip]{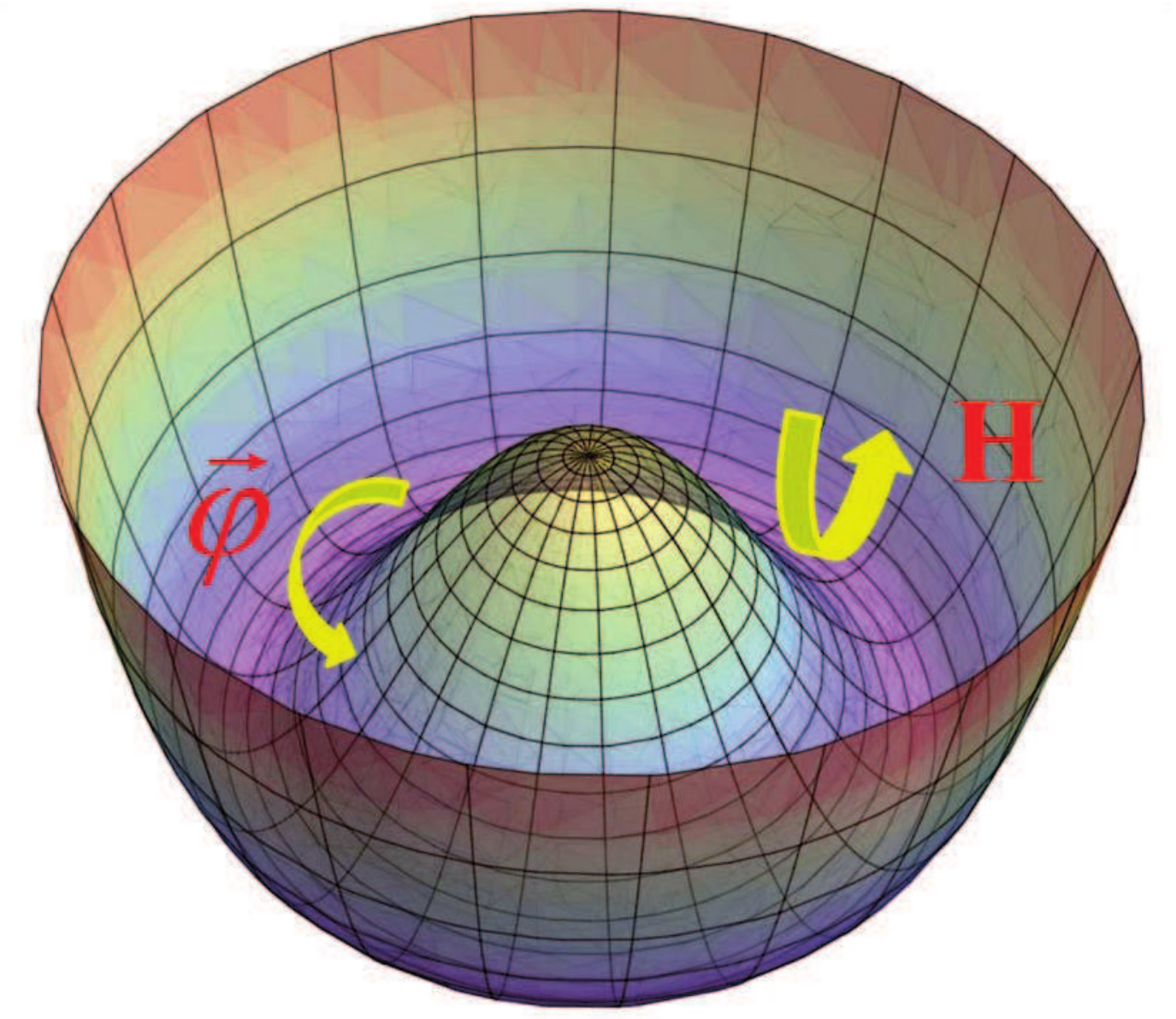}
\caption{SM scalar potential. The Goldstone ($\vec\varphi$) and Higgs ($H$) fields parameterize the directions indicated by the arrows.}
\label{fig:HiggsPotential}
\end{center}
\end{figure}
%%%%%%%%%%%%%%%%%%%%%%%%%%%%%%%%%%%%%%%%%%%%%%%%%%%%%%%%%%%%%%%%%%%

A massless (massive) spin-1 gauge boson has two (three) polarizations. To generate the missing longitudinal polarizations of the $W^\pm$ and $Z$ bosons, without breaking gauge invariance, one needs to incorporate three additional degrees of freedom. The SM adds a complex scalar doublet
\begin{equation}
\Phi(x)\; =\; \exp{\left\{\frac{i}{v}\,\vec{\sigma}\cdot\vec{\varphi}(x)\right\}}\;\frac{1}{\sqrt{2}}\,
\left[\begin{array}{c} 0\\ v+H(x)\end{array}\right]\, ,
\end{equation}
with a non-trivial potential generating the wanted EWSB:
\begin{equation}\label{eq:Lphi}
\cL_\phi\; =\; (D_\mu\Phi)^\dagger D^\mu\Phi - \lambda\, \left( |\Phi|^2 -\frac{v^2}{2}\right)^2 +\frac{\lambda}{4}\, v^4\, .
\end{equation}
In the unitary gauge, $\vec\varphi(x) = \vec{0}$, the three Goldstone fields are removed and the SM Lagrangian describes massive $W^\pm$ and $Z$ fields; their masses being generated by the derivative term in (\ref{eq:Lphi}): $M_W = M_Z \cos{\theta_W} = g v/2$. A massive scalar field $H(x)$, the Higgs, remains because $\Phi(x)$ contains a fourth degree of freedom, which is not needed for the EWSB. The scalar doublet structure provides a renormalizable model with good unitarity properties.

\begin{figure}
\centering
\includegraphics[width=7.5cm,clip]{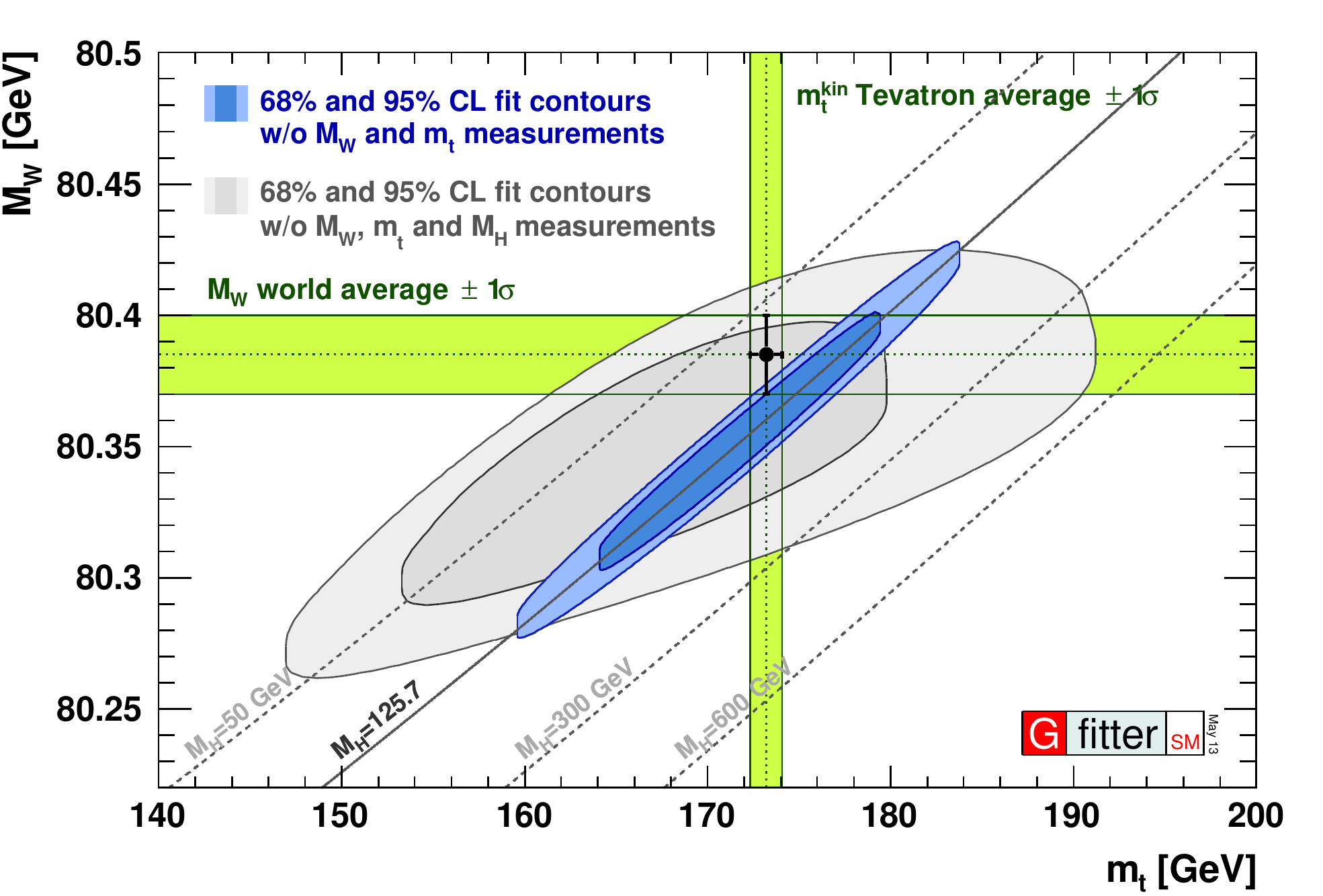}
\caption{SM electroweak fit in the $m_t$--$M_W$ plane, with (blue) and without (gray) the Higgs mass, compared with the direct measurements of the top and $W$ masses (green) \cite{Baak:2012kk}.}
\label{fig:EWfit_mt}
\end{figure}

While the vacuum expectation value (the electroweak scale) was already known,
$v = (\sqrt{2} G_F)^{-1/2} = 246$~GeV, the measured Higgs mass determines the last free parameter of the SM, the quartic scalar coupling:
\begin{equation}\label{eq:lambda}
\lambda\; =\; \frac{M_H^2}{2 v^2}\; =\; 0.13\, .
\end{equation}
As shown in figure~\ref{fig:EWfit_mt}, the measured Higgs mass is in beautiful agreement with the expectations from global fits to precision electroweak data \cite{Baak:2012kk}.

%%%%%%%%%%%%%%%%%%%%%% Figure EW fit %%%%%%%%%%%%%%%%%%%%%
%%%%%%%%%%%%%%%%%%%%%%%%%%%%%%%%%%%%%%%%%%%%%%%%%%%%%%%%%%%%%%%%%%%

\begin{figure}
\centering
\includegraphics[width=7cm,clip]{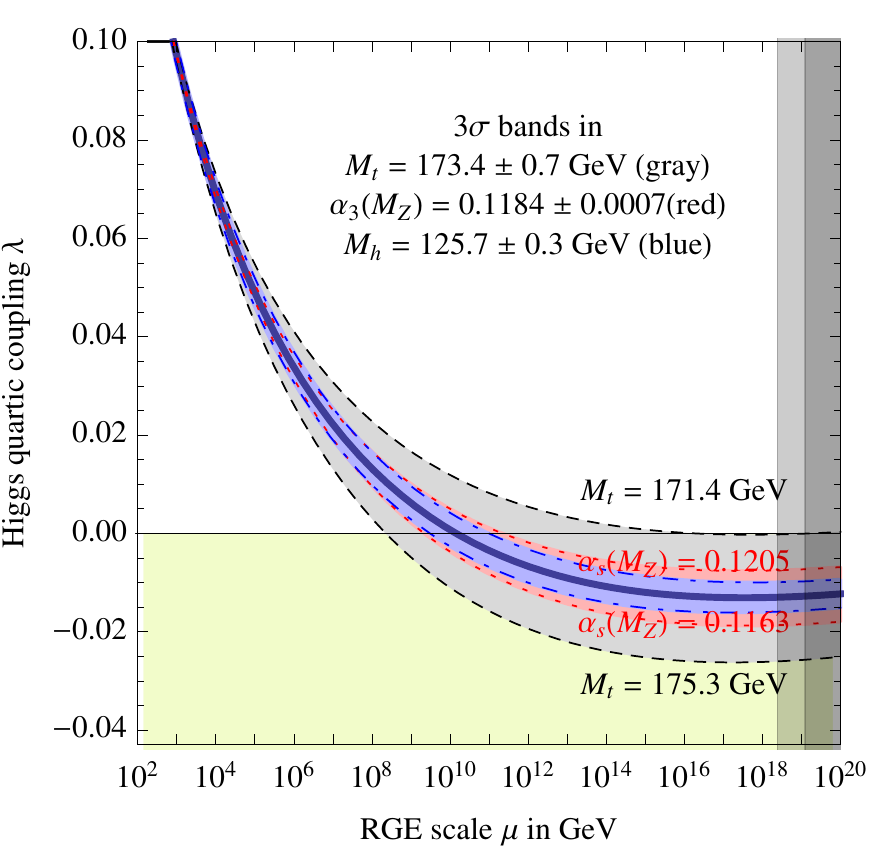}
\caption{Evolution of $\lambda(\mu)$ with the renormalization scale  \cite{Buttazzo:2013uya}.}
\label{fig:lambda}
\end{figure}

Quantum corrections to $M_H^2$ are dominated by positive contributions from heavy top loops, which grow logarithmically with the renormalization scale $\mu$. Since the physical value of $M_H$ is fixed, the tree-level contribution $2 v^2\lambda(\mu)$ decreases with increasing $\mu$. Figure~\ref{fig:lambda} shows the evolution of $\lambda(\mu)$ up to the Planck scale ($M_{\mathrm{Pl}} = 1.2\times 10^{19}$~GeV), varying $m_t$, $\alpha_s(M_Z)$ and $M_H$ by $\pm 3\sigma$ \cite{Buttazzo:2013uya}. The Higgs quartic coupling remains weak in the entire energy domain below $M_{\mathrm{Pl}}$ and crosses $\lambda=0$ at very high energies around $10^{10}$~GeV. The values of $M_H$ and $m_t$ appear to be very close to those needed for absolute stability of the potential ($\lambda >0$) up to  $M_{\mathrm{Pl}}$, which would require $M_H > (129.6\pm 1.5)$~GeV \cite{Buttazzo:2013uya,Degrassi:2012ry}
($\pm 5.6$~GeV if more conservative uncertainties on the top mass are adopted \cite{Alekhin:2012py}). Moreover, even if $\lambda$ becomes slightly negative at very high energies, the resulting potential instability leads to an electroweak vacuum lifetime much larger than any relevant astrophysical or cosmological scale. Thus, the Higgs and top masses result in a metastable vacuum \cite{Buttazzo:2013uya,Degrassi:2012ry} and the SM could be valid up to $M_{\mathrm{Pl}}$. The possibility of some new-physics threshold at scales $\Lambda\sim M_{\mathrm{Pl}}$, leading to the matching condition
$\lambda(\Lambda) = 0$, is obviously intriguing.

%%%%%%%%%%%%%%%%%%%%%% Figure EW fit %%%%%%%%%%%%%%%%%%%%%
%%%%%%%%%%%%%%%%%%%%%%%%%%%%%%%%%%%%%%%%%%%%%%%%%%%%%%%%%%%%%%%%%%%

\section{Higgs Signal Strengths}
\label{sec:data}

The data on the Higgs-like boson are conveniently expressed in terms of the so-called Higgs signal strengths, which measure the product of the Higgs production cross section times its decay branching ratio into a given final state, in units of the corresponding SM prediction:
$\mu \equiv \sigma\cdot \mathrm{Br}/(\sigma_{\mathrm{SM}}\cdot \mathrm{Br}_{\mathrm{SM}})$. Thus, the SM corresponds to $\mu=1$.
Table~\ref{tab:LHCdata} summarizes the present ATLAS \cite{Aad:2013wqa}, CMS \cite{Chatrchyan:2013lba} and Tevatron \cite{Aaltonen:2012qt} results.
The new boson appears to couple to the known gauge bosons ($W^\pm$, $Z$, $\gamma$, $G^a$)
with the strength expected for the SM Higgs. A slight excess of events ($2\sigma$) in the $2\gamma$ decay channel is observed by ATLAS, but the CMS data no-longer confirm this trend. The global LHC (world) average,
\begin{equation}
\mu\; =\; 0.96\pm 0.11
\qquad (0.98\pm 0.11)\, ,
\end{equation}
is in perfect agreement with the SM.

%%%%%%%%%%%%%%%%%%%%%% Tevatron included %%%%%%%%%%%%%%%%%%%%%%%%%%%%%%
\begin{table}
\centering
\caption{Measured Higgs Signal Strengths \cite{Aad:2013wqa,Chatrchyan:2013lba,Aaltonen:2012qt}.}
\label{tab:LHCdata}
\renewcommand{\arraystretch}{1.2} % enlarge line spacing
\begin{tabular}{lccc}
\hline
Decay Mode & ATLAS & CMS & Tevatron\\ \hline
$H\to bb$ & $0.2\,{}^{+\, 0.7}_{-\, 0.6}$ & $1.15\pm 0.62$
& $1.59\,{}^{+\, 0.69}_{-\, 0.72}$
\\
$H\to \tau\tau$ & $0.7\,{}^{+\, 0.7}_{-\, 0.6}$ & $1.10\pm 0.41$
& $1.68\,{}^{+\, 2.28}_{-\, 1.68}$
\\
$H\to \gamma\gamma$ & $1.55\,{}^{+\, 0.33}_{-\, 0.28}$ & $0.77\pm 0.27$
& $5.97\,{}^{+\, 3.39}_{-\, 3.12}$
\\
$H\to WW^*$ & $0.99\,{}^{+\, 0.31}_{-\, 0.28}$ & $0.68\pm 0.20$
& $0.94\,{}^{+\, 0.85}_{-\, 0.83}$
\\
$H\to ZZ^*$ & $1.43\,{}^{+\, 0.40}_{-\, 0.35}$ & $0.92\pm 0.28$
&
\\ \hline
Combined & $1.23\pm 0.18$ & $0.80\pm 0.14$
& $1.44\,{}^{+\, 0.59}_{-\, 0.56}$
\\\hline
\end{tabular}
\end{table}
%%%%%%%%%%%%%%%%%%%%%%%%%%%%%%%%%%%%%%%%%%%%%%%%%%%%%%%%%%%%%%%%%%%%%%%

The sensitivity to the different Higgs couplings is increased disentangling the different production channels: gluon fusion ($GG\to t\bar t\to H$), vector-boson fusion ($VV\to H$, $V=W,Z$) and associated $VH$ or $t\bar t H$ production. At the LHC, the dominant contribution comes from the gluon-fusion mechanism which gives access to the top Yukawa.
Evidence for vector-boson fusion production has been already reported with a significance above $3\sigma$.
Complementary information is provided by the Tevatron data, specially in the $VH\to Vb\bar b$ mode.

The agreement of the measured Higgs production cross section with the SM prediction confirms the existence of a top Yukawa coupling with the expected size. Moreover, it excludes the presence of additional fermionic contributions to gluon-fusion production. A fourth quark generation would increase the cross section by a factor of nine, and much larger enhancements ($\sim 4 T_R^2/T_F^2$) would result from exotic fermions in higher-colour representations, coupled to the Higgs \cite{Ilisie:2012cc}. Thus, the  present Higgs data exclude a fourth SM generation \cite{Aaltonen:2012qt,Aad:2011qi,Chatrchyan:2013sfs}, while exotic strongly-interacting fermions could only exist provided their masses are not generated by the SM Higgs \cite{Ilisie:2012cc} (or with fine-tuned cancelations with scalar loops).

%%%%%%%%%%%%%%%%%%%%%%%%% $H\to 2\gamma$ %%%%%%%%%%%%%%%%%%%%%%%%%%
\begin{figure}
\centering
\begin{minipage}{3.5cm}
\includegraphics[width=3.2cm,clip]{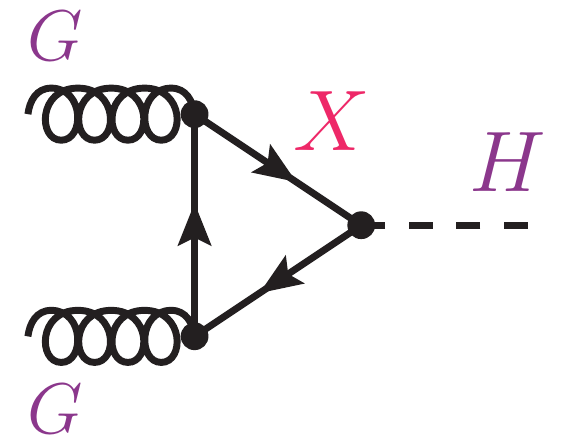}
\end{minipage}
\begin{minipage}{4cm}
$\displaystyle\sim\quad\sum_{a=1}^8 \mathrm{Tr}\left[t_R^a t_R^a\right]\; =\; 8 T_R$
\end{minipage}
\caption{$GG\to H$ production through a heavy fermion loop.}
\label{fig:ggH}
\end{figure}
%%%%%%%%%%%%%%%%%%%%%%%%%%%%%%%%%%%%%%%%%%%%%%%%%%%%%%%%%%%%%%%%%%%

%%%%%%%%%%%%%%%%%%%%%%%%% $H\to 2\gamma$ %%%%%%%%%%%%%%%%%%%%%%%%%%
\begin{figure}
\centering
\includegraphics[width=8cm,clip]{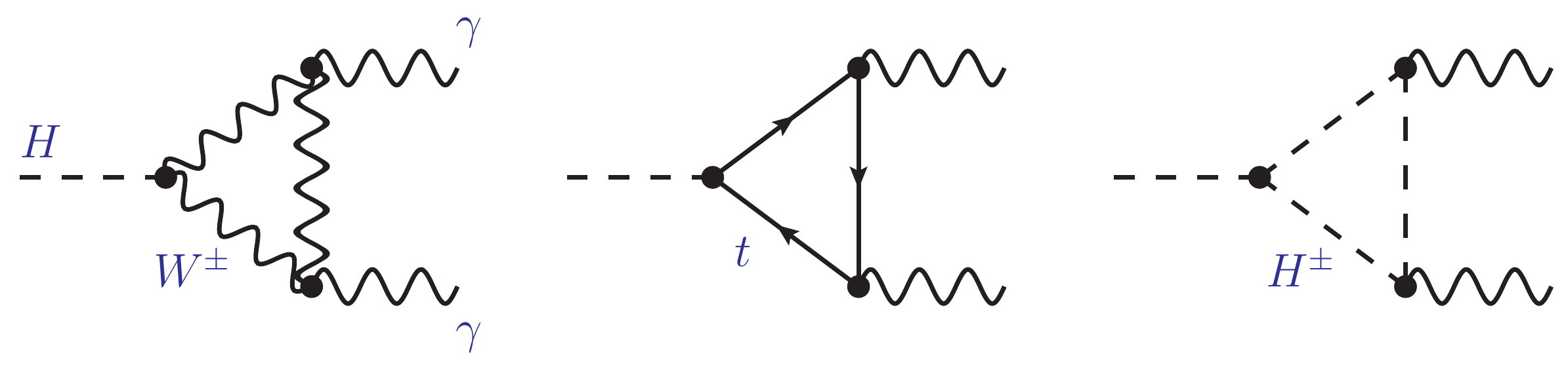}
\caption{One-loop contributions to $H\to \gamma\gamma$.}
\label{fig:H2gamma}
\end{figure}
%%%%%%%%%%%%%%%%%%%%%%%%%%%%%%%%%%%%%%%%%%%%%%%%%%%%%%%%%%%%%%%%%%%

The decay $H\to\gamma\gamma$ occurs in the SM through intermediate $W^+W^-$ and $t\bar t$
triangular loops, which interfere destructively. Therefore, it is sensitive to new physics contributions such as the charged-scalar loop shown in figure~\ref{fig:H2gamma}. The decay width roughly scales as
\begin{equation}
\Gamma(H\to\gamma\gamma)\;\sim\; \left| -8.4\, \kW + 1.8\, \kt + \cC_{\mathrm{NP}}\right|^2\, ,
\end{equation}
where $\kW$ and $\kt$ denote the $H W^+W^-$ and $H t \bar t$ couplings in SM units, and $\cC_{\mathrm{NP}}$
accounts for any additional decay amplitude beyond the SM. An enhanced rate with respect to the SM prediction ($\kW =\kt=1$, $\cC_{\mathrm{NP}}=0$) could be obtained either flipping the relative sign of the $W^\pm$ and top amplitudes ($\kW \kt <0$) or through an additional contribution with $\cC_{\mathrm{NP}}<0$. Many models have been discussed to explain the $2\gamma$ excess in this way, but the present disagreement between ATLAS and CMS does not allow to extract significant conclusions.

Taking world averages (LHC and Tevatron) for the Higgs signal strengths in the different decay and production channels, the SM provides a perfect description of the experimental data. A fit to the measured rates gives $M_H = (124.5\pm 1.7)$~GeV \cite{Giardino:2013bma}, which agrees with the value extracted from the peak position in Eq.~(\ref{eq:Higgs_Mass}).
The fermionic ($\tau$, $b$, $t$) and bosonic ($W^\pm$, $Z$) couplings of the $H$ boson seem compatible with a linear and quadratic, respectively, dependence with their masses, scaled by the electroweak scale. Fitting the data with the parameterization
\begin{equation}
\lambda_f \; =\; \sqrt{2}\;\left(\frac{m_f}{M}\right)^{1+\epsilon}\, ,
\qquad
g_{HVV}^{\phantom{2}}\; =\; 2\;\left( \frac{M_V^{2 (1+\epsilon)}}{M^{1+2\epsilon}}\right)\, ,
\end{equation}
leads to \cite{Ellis:2013lra}
\begin{equation}
\epsilon\; =\; -0.022\, {}^{+\, 0.042}_{-\, 0.021}\, ,
\qquad
M\; =\; 244\, {}^{+\, 20}_{-\, 10}\;\mathrm{GeV}\, ,
\end{equation}
in excellent agreement with the SM values $\epsilon =0$ and $M=v=246$~GeV.
Thus, $H$ has the properties expected for a Higgs-like particle, related with the EWSB.

%%%%%%%%%%%%%%%%%%%%%%%%%%%%% a/c fit %%%%%%%%%%%%%%%%%%%%%%%%%%%%%
\begin{figure}[t]
\centering
\includegraphics[width=7cm,clip]{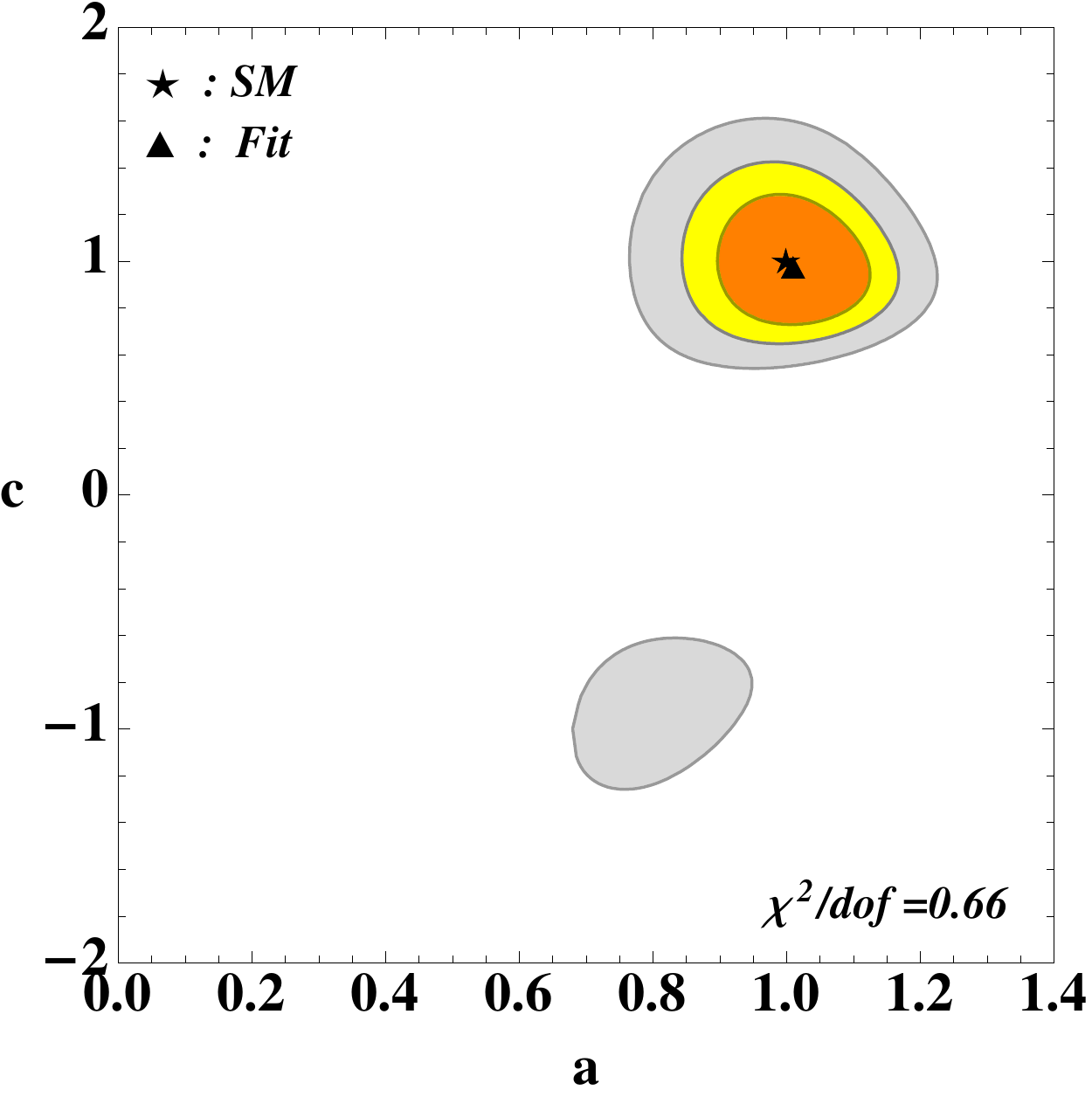}
\vskip -.2cm
\caption{Two-parameter fit to the Higgs signal strengths,
at 68\% (orange), 90\% (yellow) and 99\% (gray) CL \cite{CIP:13}.}
\label{fig:a_c_fit}
\end{figure}
%%%%%%%%%%%%%%%%%%%%%%%%%%%%%%%%%%%%%%%%%%%%%%%%%%%%%%%%%%%%%%%%%%%

Figure~\ref{fig:a_c_fit} \cite{CIP:13} shows an equivalent two-parameter fit in terms of the more usual
$a=\kV$ and $c=\kF$ parameterization of the bosonic and fermionic couplings in SM units; {\it i.e.}, the $HW^+W^-$ and $HZZ$ couplings of the SM are rescaled with a common factor $a$ and all fermionic Yukawas are multiplied by $c$. One obtains \cite{CIP:13},
\begin{equation}\label{eq:a-c-fit}
a\; =\; 1.01\pm 0.07\, ,
\qquad
c\; =\; 0.97\, {}^{+\, 0.20}_{-\, 0.17}\, .
\end{equation}
The SM point $a=c=1$ is right at the center of the 68\% CL region. A second solution with a flipped Yukawa sign appears only at 99\% CL; the confidence level of this solution increases to 68\% if only the ATLAS data is included in the fit, owing to the enhanced $H\to\gamma\gamma$ rate.

\section{Two-Higgs Doublet Models}
\label{sec:A2HDM}

Two-Higgs-doublet models provide a minimal extension of the SM scalar sector that naturally accommodates the electroweak precision tests.
The enlarged scalar spectrum contains one charged ($H^\pm$) and three neutral ($\varphi_i^0 =\left\{ h, H, A\right\}$) Higgs bosons. In full generality, one can choose a basis in the scalar space so that only the first doublet acquires a vacuum expectation value, playing the role of the SM scalar doublet with one Higgs-like CP-even neutral field ($S_1$) and the 3 Goldstones. The second doublet contains the $H^\pm$ boson and two neutral fields, one of them CP-even ($S_2$) and the other CP-odd ($S_3$).
The neutral mass eigenstates $\varphi_i^0 = \cR_{ij} S_j$ are defined through an orthogonal rotation matrix $\cR$, which is determined by the scalar potential. In the limit of CP conservation, $\cR$ reduces to a dimension-2 rotation of angle $\tilde\alpha$, mixing the CP-even states $S_1$ and $S_2$, and $A=S_3$.

Since $g_{\varphi^0_i VV} = \cR_{i1}\, g_{HVV}^{\mathrm{SM}}$, the strength of the SM gauge coupling is shared by the three neutral scalars:
\begin{equation}
g^2_{hVV} +  g^2_{HVV} + g^2_{AVV} \; =\; \left(g_{HVV}^{\mathrm{SM}}\right)^2\, .
\end{equation}
Therefore, the gauge coupling of each scalar is predicted to be smaller than the SM one
($g_{AVV}=0$ if CP is conserved).

Generic multi-Higgs doublet models give rise to unwanted flavour-changing neutral-current (FCNC) interactions, through non-diagonal couplings of neutral scalars to fermions.
The tree-level FCNCs can be eliminated, requiring the alignment in flavour space of the two Yukawa matrices coupling to a given right-handed fermion state. The Yukawa couplings of the aligned two-Higgs-doublet model (A2HDM) \cite{Pich:2009sp} are fully characterized by the three complex alignment parameters $\varsigma_f^{\phantom{.}}$ ($f=u,d,\ell$), which provide new sources of CP violation. For particular (real) values of $\varsigma_f^{\phantom{.}}$, one recovers the usual models with natural flavour conservation, based on discrete $Z_2$ symmetries.

%%%%%%%%%%%%%%%%%%%%%%%%%%% eta_f fit %%%%%%%%%%%%%%%%%%%%%%%%%%%%%
\begin{figure}[t]
\centering
\includegraphics[width=7.7cm,clip]{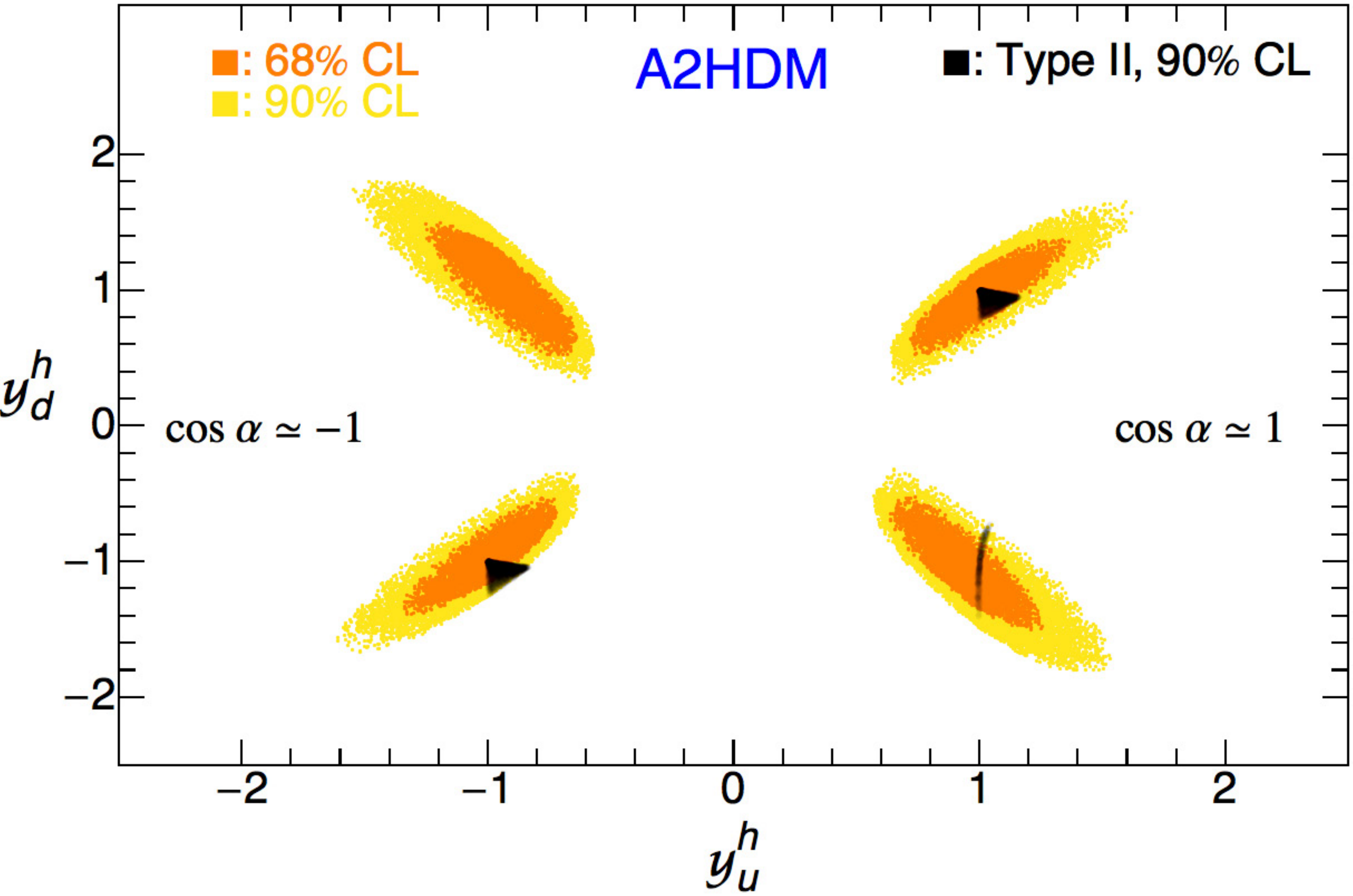}
\caption{Allowed 68\% and 90\% CL regions in the $y_u^h$--$y_d^h$ plane
\cite{CIP:13,Celis:2013rcs}, within the CP-conserving A2HDM. The small black areas correspond to the usual type II model at 90\% CL.}
\label{fig:eta_fit}
\end{figure}
%%%%%%%%%%%%%%%%%%%%%%%%%%%%%%%%%%%%%%%%%%%%%%%%%%%%%%%%%%%%%%%%%%%

Let us assume the 126 GeV boson to be the lightest neutral state $h$. Neglecting CP-violation effects, its fermionic and gauge couplings are given, in SM Higgs units, by
\begin{equation}
\kf\; =\; y_f^h\; =\; \cos{\tilde\alpha} + \varsigma_f^{\phantom{.}}\,\sin{\tilde\alpha}\, ,
\qquad
\kV\; =\; \cos{\tilde\alpha}\, .
\end{equation}
A global fit to the Higgs data \cite{CIP:13,Celis:2013rcs}, obeying the known flavour constraints \cite{Jung:2010ik},
results in four possible regions in the $(y_u^h,y_d^h,y_\ell^h)$ space, which in figure~\ref{fig:eta_fit} are projected into the plane $y_u^h$--$y_d^h$.
Since the $H^\pm$ contribution to $h\to 2\gamma$ has been assumed to be negligible, the fit requires the top and $W^\pm$ amplitudes to have the same relative sign as in the SM; {\it i.e.}, $\kt\equiv\ku$ and $\kV$ are forced to have identical signs. Moreover, $|\cos{\tilde\alpha}|> 0.80$ at 90\% CL. The figure shows also how the allowed regions shrink in the particular case of the type II model ($\varsigma_{d,\ell}^{\phantom{.}} = -1/\varsigma_u^{\phantom{.}} = -\tan{\beta}$), usually assumed in the literature and realized in minimal supersymmetric scenarios. This clearly illustrates that there is a much wider range of open phenomenological possibilities waiting to be explored.
The couplings of the missing Higgs bosons $H^\pm$, $H$ and $A$, and therefore their phenomenology, are very different in each of the allowed regions shown in figure~\ref{fig:eta_fit} \cite{CIP:13,Celis:2013rcs}.

\section{Custodial Symmetry}

It is convenient to collect the SM Higgs doublet $\Phi$ and its charge-conjugate
$\Phi^c = i\sigma_2 \Phi^*$ into the $2\times 2$ matrix
\begin{equation}\label{eq:Sigma}
\Sigma\,\equiv\,\left(\Phi^c , \Phi\right)\, =\,
\left(\begin{array}{cc} \Phi^{0*} & \Phi^+ \\ -\Phi^- & \Phi^0\end{array}\right)
\, =\, \frac{1}{\sqrt{2}}\; (v+H)\; U(\vec\varphi)\, ,
\end{equation}
where the 3 Goldstone bosons are parameterized through
\begin{equation}\label{eq:Uphi}
U(\vec\varphi)\;\equiv\;
\exp{\left\{\frac{i}{v}\:\vec{\sigma}\cdot\vec{\varphi}(x)\right\}}\, .
\end{equation}
Dropping the constant term $\lambda v^4/4$, the SM scalar Lagrangian (\ref{eq:Lphi})
takes the form
\begin{eqnarray}\label{eq:Lsigma}\hskip -.2cm
\cL_\phi &\!\! =&\!\! \frac{1}{2}\;\mathrm{Tr} \left[ (D_\mu\Sigma)^\dagger D^\mu\Sigma\right]
- \frac{\lambda}{16}\,\left(\mathrm{Tr} \left[ \Sigma^\dagger \Sigma\right] - v^2\right)^2
\nonumber\\
&\!\! =&\!\! \frac{v^2}{4}\;\mathrm{Tr} \left[ (D_\mu U)^\dagger D^\mu U\right]
\; +\; \cO(H/v)\, .
\end{eqnarray}
This expression makes manifest the existence of a global
$\mathrm{SU(2)}_L\otimes \mathrm{SU(2)}_R$ symmetry ($\Sigma\to g_L^{\phantom{\dagger}}\,\Sigma\, g_R^\dagger$,
$g_X^{\phantom{\dagger}}\in\mathrm{SU(2)}_X$)
which is broken by the vacuum to the diagonal $\mathrm{SU(2)}_{L+R}$.
The SM promotes the $\mathrm{SU(2)}_L$ to a local gauge symmetry, while only the $\mathrm{U(1)}_Y$ subgroup of $\mathrm{SU(2)}_R$ is gauged; thus, the $\mathrm{SU(2)}_R$ symmetry is explicitly broken at $\cO(g')$ through the $\mathrm{U(1)}_Y$ interaction in the covariant derivative.
The second line in (\ref{eq:Lsigma}), without the Higgs field, is the generic Goldstone Lagrangian associated with this type of symmetry breaking, which is responsible for the successful generation of the $W^\pm$ and $Z$ masses. The same Lagrangian describes the low-energy chiral dynamics of the QCD pions, with the notational changes $v\to f_\pi$ and $\vec\varphi\to\vec\pi$.

\section{Strongly-Coupled Scenarios}
\label{sec:strong}

The recently discovered boson could be a first experimental signal of a new strongly-interacting sector: the lightest state of a large variety of new resonances
of different types as happens in QCD. Among the many possibilities (technicolour, walking technicolour, conformal technicolour, higher dimensions \ldots), the relatively light mass of the discovered Higgs candidate has boosted the interest on strongly-coupled scenarios with a composite pseudo-Goldstone Higgs boson \cite{Espinosa:2010vn},
where the Higgs mass is protected by an approximate global symmetry and is only generated via quantum effects. A simple example is provided by the popular $\mathrm{SO(5)}/\mathrm{SO(4)}$ minimal composite Higgs model \cite{Agashe:2004rs,Contino:2006qr}.
Another possibility would be to interpret the Higgs-like scalar as a dilaton, the pseudo-Goldstone boson associated with the spontaneous breaking of scale invariance at some scale $f_\varphi\gg v$ \cite{Goldberger:2007zk,Matsuzaki:2012xx,Bellazzini:2012vz,Chacko:2012vm}.

The dynamics of Goldstones and massive resonances can be analyzed in a model-independent way by using a low-energy effective Lagrangian based on the known pattern of EWSB, $\mathrm{SU(2)}_L\otimes \mathrm{SU(2)}_R\to\mathrm{SU(2)}_{L+R}$.
To lowest order in derivatives and number of resonance fields~\cite{Pich:2012dv},
\begin{eqnarray}\label{eq:EWET} \hskip -.2cm
\mathcal{L} &\!\! = &\!\!
\frac{v^2}{4}\;\mathrm{Tr} \left[ (D_\mu U)^\dagger D^\mu U\right]\;
\left( 1 + \frac{2\,\omega}{v}\, H \right)
\\ &\!\! + &\!\!
\frac{F_A}{2\sqrt{2}}\, \mathrm{Tr} \left[ A_{\mu\nu} f^{\mu\nu}_- \right]
\, +\, \frac{F_V}{2\sqrt{2}}\, \mathrm{Tr} \left[ V_{\mu\nu} f^{\mu\nu}_+ \right]
\, +\, \cdots
\nonumber
\end{eqnarray}
The first term gives the (gauged) Goldstone Lagrangian, plus their interactions
with the $\mathrm{SU(2)}_{L+R}$ singlet Higgs-like particle $H$.
For $\omega=1$ one recovers the $H\varphi\varphi$ vertex of the SM; {\it i.e.},
the SM Higgs coupling to the gauge bosons ($\omega = a = \kV$).
The effective Lagrangian also incorporates the lightest vector and axial-vector resonance multiplets $V_{\mu\nu}$ and $A_{\mu\nu}$ with masses $M_V$ and $M_A$.
The $F_V$ and $F_A$ terms couple these resonances with the gauge and Goldstone fields through $f^{\mu\nu}_\pm$.

We have already seen in Eq.~(\ref{eq:a-c-fit}) that the LHC data requires $\omega$ to be within 10\% of its SM value. A much stronger constraint is obtained from the measured $Z$ and $W^\pm$ self-energies \cite{Pich:2012dv,Espinosa:2012ir,Falkowski:2013dza}, which are modified by the presence of massive resonance states coupled to the gauge bosons. The effect is characterized by the so-called oblique parameters \cite{Peskin:1990zt};
the global fit to electroweak precision data determines the values
$S = 0.03\pm 0.10$ and $T=0.05\pm0.12$~\cite{Baak:2012kk}.
$S$ receives tree-level contributions from vector and axial-vector exchanges, while
$T$ is identically zero at lowest-order (it measures the breaking of custodial symmetry).

Imposing a good short-distance behaviour of the effective theory,\footnote{
%%%%%%%%%%%%%%%% Footnote
One requires the validity of the two Weinberg sum rules \cite{Weinberg:1967kj}, which are known to be true in asymptotically-free gauge theories. %  \cite{Bernard:1975cd}
The results are slightly softened if one only imposes the first sum rule, which
is also valid in gauge theories with non-trivial ultraviolet fixed points.
}
%%%%%%%%%%%%%%%%%%%%%%%%%
the tree-level contribution to $S$ is determined by the resonance masses. The experimental constraint on $S$
implies that $M_{V,A}$ are larger than 1.8 (2.4) TeV at 95\% (68\%) CL. Thus, strongly-coupled models of EWSB should have a quite high dynamical mass scale. While this was often considered to be an undesirable property, it fits very well with the LHC findings which are pushing the scale of new physics beyond the TeV region. It also justifies our approximation of only considering the lightest resonance multiplets.
The NLO contributions to $S$ from $\varphi\varphi$, $V\varphi$ and $A\varphi$ loops are small and make slightly stronger the lower bound on the resonance mass scale \cite{Pich:2012dv}.

%%%%%%%%%%%%%%%%%%%%%% Oblique constraint %%%%%%%%%%%%%%%%%%%%%%%%
\begin{figure}[t]
\centering
\includegraphics[width=6.7cm,clip]{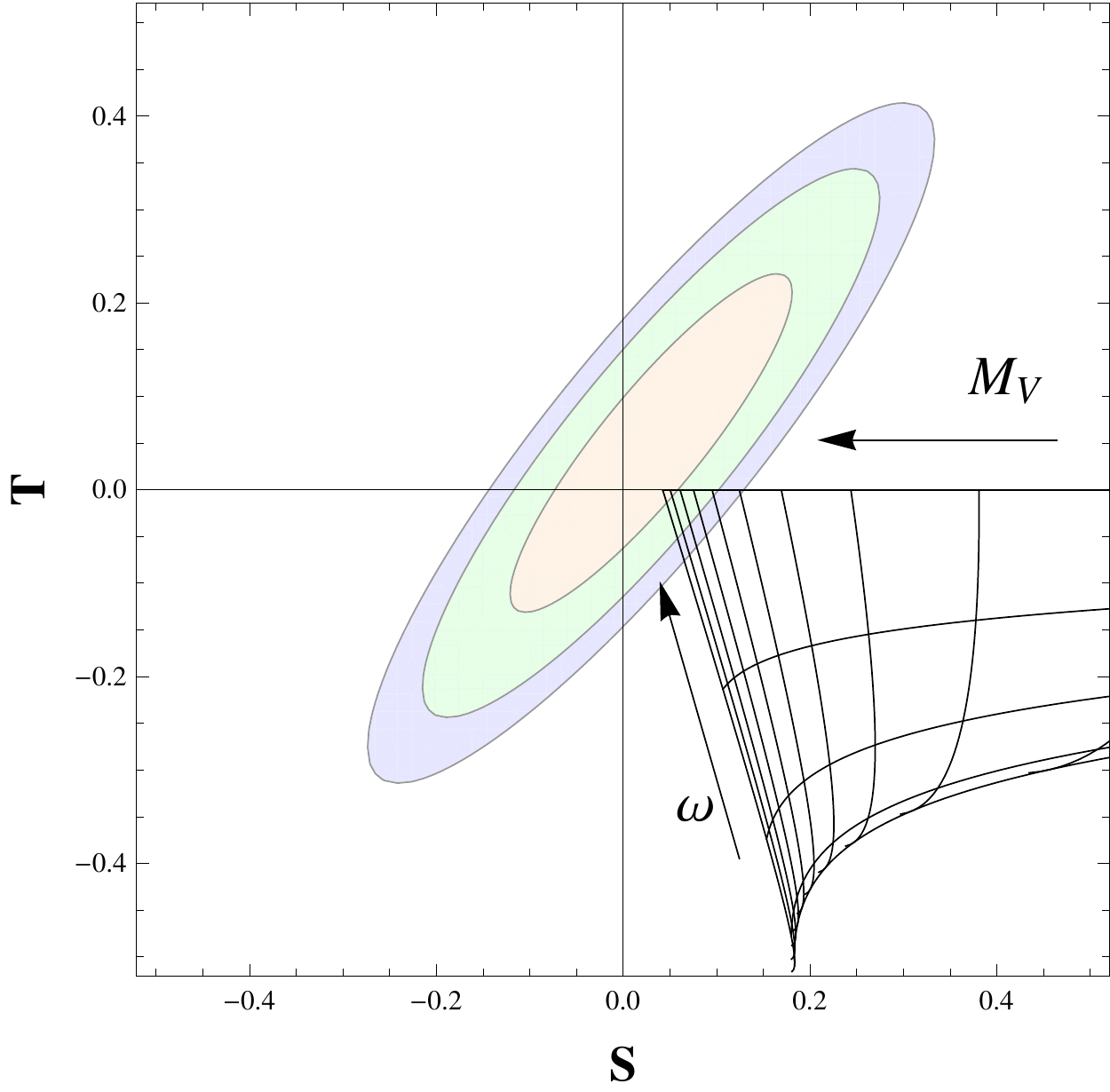}
\caption{NLO determination of $S$ and $T$.
The grid lines correspond to $M_V$ values from $1.5$ to $6.0$~TeV, at intervals of $0.5$~TeV, and $\omega= 0, \, 0.25, 0.5, 0.75, 1$.
The arrows indicate the directions of growing  $M_V$ and $\omega$.
The ellipses give the experimentally allowed regions at 68\%, 95\% and 99\% CL~\cite{Pich:2012dv}.}
\label{fig:ST-2WSR}
\end{figure}
%%%%%%%%%%%%%%%%%%%%%%%%%%%%%%%%%%%%%%%%%%%%%%%%%%%%%%%%%%%%%%%%%%%

Much more important is the presence of a light scalar resonance with $M_H= 126$ GeV. Although it does not contribute at LO, there exist sizeable $H B$ ($H\varphi$) loop contributions to $T$ ($S$), which are proportional to $\omega^2$
($B$ is the $\mathrm{U}(1)_Y$ gauge field).
Figure~\ref{fig:ST-2WSR}
compares the NLO theoretical predictions with the experimental bounds \cite{Pich:2012dv}.
At 68\% (95\%) CL, one gets $\omega\in [0.97,1]$  ($[0.94,1]$),
in nice agreement with the present LHC evidence but much more restrictive.
Moreover, the vector and axial-vector states should be very heavy (and quite degenerate);
one finds $M_A \approx M_V> 5$~TeV ($4$~TeV) at 68\% (95\%) CL \cite{Pich:2012dv}.

These conclusions are quite generic, since only rely on mild assumptions about the ultraviolet behaviour of the underlying strongly-coupled theory, and can be easily particularized to more specific models.
The dilaton coupling to the electroweak bosons corresponds to $\omega = v/f_\varphi$, which makes this scenario quite unlikely. More plausible could be the pseudo-Goldstone Higgs identification. In the $\mathrm{SO(5)}/\mathrm{SO(4)}$ minimal models,
$\omega = (1-v^2/f_\varphi^2)^{1/2}$ \cite{Agashe:2004rs,Contino:2006qr} with $f_\varphi$ the typical scale of the Goldstone bosons of the strong sector, which is then tightly constrained by electroweak (and LHC) data.

Thus, strongly-coupled electroweak models are allowed by current data provided the resonance mass scale stays above the TeV and the light Higgs-like boson has a gauge coupling close to the SM one. This has obvious implications for future LHC studies, since it leads to a SM-like scenario. A possible way out would be the existence of additional light scalar degrees of freedom, sharing the strength of the SM gauge coupling as happens in (weakly-coupled) two-Higgs-doublet models.

Values of $\omega\not= 1$ lead to tree-level unitarity violations in the scattering of two longitudinal gauge bosons. The present experimental constraints on $\omega$ imply already that the perturbative unitarity bound is only reached at very high energies above 3 TeV.
Unitarity violations could also originate from anomalous gauge self-interactions, which are
again bounded by collider data \cite{Corbett:2013pja}. A recent study of the implications of unitarity in the strongly-interacting electroweak context has been given in Ref.~\cite{Espriu:2012ih}.

\section{Discussion}

The successful discovery of a boson state at the LHC brings a renewed perspective in particle physics. The new boson behaves indeed as the SM Higgs and its mass fits very well with the expectations from global fits to precision electroweak data.
Thus, the SM appears to be the right theory at the electroweak scale and all its parameters and fields have been finally determined. In fact, with the measured Higgs and top masses, the SM could be a valid theory up to the Planck scale.

However, new physics is still needed to explain many pending questions for which the SM does not provide satisfactory answers. A proper understanding of the vastly different mass scales spanned by the known particles is missing. The dynamics of flavour and the origin of CP violation are also related to the mass generation.
The Higgs-like boson could be a window into unknown dynamical territory. Thus, its properties must be analyzed with high precision to uncover any possible deviation from the SM. The present data are already putting stringent constraints on alternative scenarios of EWSB and pushing the scale of new physics to higher energies. How far this scale could be is an open question of obvious experimental relevance.

If new physics exits at some scale $\Lambda_{\mathrm{NP}}$, quantum corrections of the type $\delta M_H^2\sim g^2/(4\pi)^2 \Lambda_{\mathrm{NP}}^2\log{(\Lambda_{\mathrm{NP}}^2/M_H^2)}$ could bring $M_H$ to the heavy scale $\Lambda_{\mathrm{NP}}$.
Which symmetry keeps $M_H$ away from $\Lambda_{\mathrm{NP}}$? Fermion masses are protected by chiral symmetry, while gauge symmetry protects the gauge boson masses; those particles are massless when the symmetry becomes exact.
Supersymmetry was originally advocated to protect the Higgs mass, but according to present data this no-longer works `naturally'. Another possibility would be scale symmetry, which is broken by the Higgs mass; a naive dilaton is basically ruled out, but there could be an underlying conformal theory at $\Lambda_{\mathrm{NP}}$. Dynamical EWSB with light pseudo-Goldstone particles at low energies remains also a viable scenario.
Future discoveries at the LHC should bring a better understanding of the correct dynamics above the electroweak scale.

\section*{Acknowledgments}

I would like to thank A. Celis and V. Ilisie for their help and useful comments.
This work has been supported in part by the Spanish Government and EU funds for regional development [grants FPA2007-60323, FPA2011-23778 and CSD2007-00042 (Consolider Project CPAN)], and the Generalitat Valenciana [PrometeoII/2013/007].

\end{document}